\documentstyle[prc,aps,epsf]{revtex}
\newcommand{\beq}{\begin{equation}}
\newcommand{\eeq}{\end{equation}}
\newcommand{\beqa}{\begin{eqnarray}}
\newcommand{\eeqa}{\end{eqnarray}}
\begin{document}
\draft
\preprint{MKPH-T-99-20}
\title{%
Elastic electron deuteron scattering with consistent meson exchange and 
relativistic contributions of leading order\thanks{Supported
by the Deutsche Forschungsgemeinschaft (SFB 443)}}
\author{
Hartmuth Arenh{\"o}vel, Frank Ritz and Thomas Wilbois\footnote{Present 
address: debis Systemhaus, Magirusstrasse 43, D-89079 Ulm, Germany}
}
\address
{Institut f{\"u}r Kernphysik, Johannes Gutenberg-Universit{\"a}t, 
D-55099 Mainz, Germany}
\date{\today}
\maketitle
\begin{abstract}
  The influence of relativistic contributions to elastic electron 
  deuteron scattering is studied systematically at low and intermediate 
  momentum transfers ($Q^2\leq 30$ fm$^{-2}$). 
  In a $(p/M)$-expansion, all leading order
  relativistic $\pi$-exchange contributions consistent with the Bonn
  OBEPQ models are included.  In addition, static heavy meson exchange
  currents including boost terms and lowest order $\rho\pi\gamma$-currents 
  are considered.  Sizeable effects from
  the various relativistic two-body contributions, mainly from
  $\pi$-exchange, have been found in form factors, structure functions 
  and the tensor polarization $T_{20}$.
  Furthermore, static properties, viz.\ magnetic dipole and charge 
  quadrupole moments and the mean square charge radius are evaluated. 
\end{abstract}


\section{Introduction}
Recently, we had studied systematically for photo- and electrodisintegration 
of the deuteron the influence of relativistic contributions of leading 
order in a $p/M$-expansion \cite{RiG97,RiA98}. In order to have a 
consistent framework of all one- and two-body current and boost 
contributions, we had chosen as interaction model the various Bonn OBEPQ 
versions 
\cite{MHE87,Mac89}. In particular, we were interested in the role of 
heavy meson exchange. As general result we found that in 
electrodisintegration the $\rho$ meson gives the most important 
heavy meson contribution whereas the influence of $\eta$, $\omega$, $\sigma$,
$\delta$, and $(\rho/\omega)\pi\gamma$ is much smaller, in some observables
completely negligible, in particular, near the quasifree kinematics. 
These findings were also valid for photodisintegration with some 
modifications, for example, in contrast to electrodisintegration the boost 
effects were almost negligible in photodisintegration for the energies 
considered because of the much smaller momentum transfers involved.

As a further step in these investigations it is naturally to study elastic 
electron scattering off the deuteron. On the one hand one would expect 
larger interaction effects, because also the nucleons in the final state 
are always off-shell, on 
the other hand, the leading order nonrelativistic meson exchange currents 
(MEC) from pion exchange will be absent due to their isovector character. 
Thus relativistic contributions to the MEC are expected to be more important. 
For this reason it appears mandatory to include all leading order 
relativistic contributions from pion exchange to one- and two-body 
charge and current densities including also wave function boost terms
in a consistent framework. This is the main motivation for the present work 
in which we have used the same theoretical approach than in our  
previous investigations of deuteron photo- and electrodisintegration for the 
evaluation of the invariant form factors for elastic electron deuteron 
scattering. 

Quite a few studies of this process exist in the literature which can be 
divided into two classes: (i) nonrelativistic approaches with relativistic 
contributions of leading order included, and (ii) covariant approaches. 
For a review of the status of the latter approaches see, e.g., more recent 
surveys in \cite{Gro91,Wal98} and references therein. Furthermore, we would 
like to mention the most recent work by Phillips 
{\it et al.}~\cite{PhW99} using 
a genuine three-dimensional relativistic framework.
With respect to the first class of approaches, 
very few have adopted a consistent framework. Early developments may be 
found in \cite{Gou66} and in the reviews \cite{Lev74,Cio80}. More recently, 
relativistic two-body currents from static pion and
heavy meson exchange have been studied in 
\cite{MoR89,BuY89,ScR91,TaN92,AdG93,WiS95,HeA95,PlC95}.
Mosconi and Ricci~\cite{MoR89} have studied the dependence of $A(Q^2)$ 
on the parametrization of the elementary nucleon form factors in the region
$Q^2\leq 20$~fm$^{-2}$. Their calculation is based on the Paris potential, 
and as current contributions they have included leading order relativistic
one-body terms with boost, $\pi$- and $\rho$-exchange and 
$\rho\pi\gamma$-current, but no other heavier mesons. Within a quark model 
approach for the $NN$-interaction and the MEC, Buchman 
{\it et al.}~\cite{BuY89} 
have studied this process with inclusion of leading order relativistic 
contributions to the currents, but without boost of the wave functions. 
Schiavilla and Riska~\cite{ScR91} have calculated form factors and 
observables using a current operator constructed consistently with the 
Argonne $v_{14}$ potential including relativistic contributions, but again
boost contributions have been left out completely. The same approach has 
been used by Wiringa {\it et al.}~\cite{WiS95} for the newly developed 
charge-independence breaking Argonne $v_{18}$ potential. In this work also 
static properties of the deuteron are reported. 
Within a pure one-pion-exchange model, the role of 
unitary equivalence of relativistic contributions to the charge density 
operator has been studied by Adam {\it et al.}~\cite{AdG93} using a consistent 
approach for all leading order contributions to the charge density 
operator including boost terms. Essentially the same approach but using the 
realistic Paris and Bonn OBEPQ-B potentials has been applied to the charge 
and quadrupole form factors by Henning {\it et al.}~\cite{HeA95}, but the 
magnetic form factor has not been considered. 
Plessas {\it et al.}~\cite{PlC95} have studied the influence of different 
parametrizations of the nucleon form factors on the observables. For the 
realistic Nijmegen and various Bonn potential models, they have included 
the pion pair and retardation currents, the usual relativistic one-body 
currents, and probably also the one-body boost. But the contributions from 
heavier meson exchange have been left out except for the 
$\rho\pi\gamma$-current. 

The most extensive treatment of all leading order terms including boost has 
been presented by Tamura {\it et al.}~\cite{TaN92} for elastic and 
inelastic electron deuteron scattering
based on a one-boson-exchange (OBE) model for the $NN$-interaction which 
they had constructed specifically for this purpose and it is difficult 
to assess the general quality of this potential. In addition, 
they have also considered as a realistic $NN$-potential the 
Paris potential. Since this is a phenomenological potential and not a 
genuine OBE-potential, they had taken empirical values for meson-nucleon 
coupling constants and cut-off parameters. In view of the fact, that probably 
their 
OBEP has not reached the sophistication of the realistic Bonn OBEPQ models, 
the present work appears appropriate and justified in evaluating elastic 
electron deuteron scattering within a consistent approach based on a 
realistic genuine OBE-potential. 

In the next section we will give a brief review of the relevant formalism 
for elastic electron deuteron scattering where the definition of 
form factors and structure functions are given. In Section~\ref{results} 
we first will sketch shortly the calculational framework on which our 
evaluation is based. Then the results on static electromagnetic 
properties, on form factors and structure functions will be presented and 
discussed. 

\section{Brief review of formalism}
\label{formalism}
We will start with the general expression for elastic electron scattering 
cross section off a deuteron in the laboratory system in the 
one-photon-exchange approximation for unpolarized beam and target 
\beq
\frac{d\sigma^{lab}}{d\Omega_e^{lab}}=\sigma_{Mott}^{lab}
\Big[(1+\eta)^{-2}\frac{\vec q_{lab}^{\,2}}{\vec q_{c}^{\,2}}
\,W_{L}^{(c)}(Q^2)
+\Big(\frac{1}{2}(1+\eta)^{-1}+\tan^2\frac{\theta_e^{lab}}{2}\Big)\,
W_{T}^{(c)}(Q^2)\Big]\,,\label{diffcross}
\eeq
where 
\beq
\sigma_{Mott}^{lab}=
\frac{\alpha^2\,\cos^2\frac{\theta_e^{lab}}{2}}{4
\sin^4\frac{\theta_e^{lab}}{2}}\,\frac{k'_{lab}}{k_{lab}^3}
\eeq 
denotes the Mott cross section in the laboratory system with initial and final 
electron four momenta $k_\mu$ and $k'_\mu$, respectively, $\alpha$ the fine 
structure constant, and the Lorentz scalar $\eta$ is given by
\beq
\eta=\frac{Q^2}{4M_d^2}\,.
\eeq
Here we have defined $Q^2=-q_\mu^2$, where 
$q_\mu=k_\mu-k_\mu'=(\omega,\vec q\,)$ denotes the four 
momentum transfer with 
$q_\mu^2<0$ and $M_d$ the deuteron mass. We further note the following 
relations, expressing the lab energy and three-momentum transfers by $\eta$, 
\beq
\vec q_{lab}^{\,2} = 4M_d^2\eta(1+\eta)\,,\qquad q_{0,\,lab}=2M_d\,\eta\,.
\eeq

The longitudinal and transverse 
structure functions $W_{L}^{(c)}$ and $W_{T}^{(c)}$, respectively, are 
related to the deuteron current matrix elements by
\beqa
W_{L}^{(c)}(Q^2)&=&\frac{1}{12M_d^2}
\sum_{m' m}|\langle d' m'|J_0(0)|d\, m\rangle^{(c)}|^2\,,\label{wlong}\\
W_{T}^{(c)}(Q^2)&=&\frac{1}{12M_d^2}\sum_{\lambda=\pm 1}
\sum_{m' m}|\langle d' m'|J_\lambda(0)|d\, m\rangle^{(c)}|^2\,,\label{wtrans}
\eeqa
where $d$ and $d'$ denote the four momenta of initial and final 
deuteron states which are covariantly normalized as
\beq
\langle d' m'|dm\rangle=(2\pi)^3 2 E_d \delta_{m'm}
\delta(\vec d^{\,\prime}-\vec d\,)\,,
\eeq
where $E_d=\sqrt{M_d^2+\vec d^{\,2}}$. 
Here, $J_{\pm 1}$ denote the components of the current density operator 
transverse to the momentum transfer $\vec q$, and $J_0$ the following 
combination of charge density and longitudinal current density component
\beq
J_0 = -\frac{|{\vec q}\,|^2}{q_\mu^2}
(\rho - \frac{\omega}{|{\vec q}\,|^2}
{\vec q} \cdot {\vec J}\,)
= \rho - \frac{\omega }{q_\mu^2}(\omega\rho - {\vec q} 
\cdot {\vec J}\,)\,,
\eeq
which reduces to the charge density  $\rho$ for a conserved current. 

The superscript ``$(c)$'' in (\ref{wlong}) and (\ref{wtrans})
indicates that the matrix elements and thus the structure functions in 
(\ref{diffcross}) are evaluated in a frame of reference ``c'' collinear 
to $\vec q$. Usually 
one chooses either the laboratory, the Breit or the antilab system. 
While $W_{T}^{(c)}(Q^2)$ is invariant for boosts collinear to the momentum 
transfer, the boost 
transformation of $W_{L}^{(c)}(Q^2)$ from the system ``c'' to the lab system 
is taken care of in (\ref{diffcross}) by the factor 
$\vec q_{lab}^{\,2}/\vec q_{c}^{\,2}$. Here, $\vec q_{c}$ denotes the 
three-momentum transfer in the $c$-system. The boost property of $W_{L}^{(c)}$ 
arises from the fact that $\frac{J_0}{|\vec q|}$ is invariant under 
collinear boosts. 
We note in passing that for elastic scattering the 
lab and antilab systems are equivalent for the evaluation since in this 
case one has $\vec q_{lab}=\vec q_{antilab}$.

Now we switch to noncovariant normalization and eliminate the 
c.m.\ motion by introducing the internal deuteron rest frame wave function
$|1m\rangle$ writing
\beq
|d\, m\rangle = \sqrt{2E_d}\,|\vec d\,\rangle U(\vec d\,)|1\,m\rangle\,.
\eeq
Furthermore,  
$|\vec d\,\rangle$ denotes the plane wave of the c.m.\ motion and 
$U(\vec d\,)$ the unitary operator of the boost from the deuteron 
rest system to the moving frame. The current matrix element is then reduced to 
the evaluation of the Fourier component $\tilde J_\lambda$ of the current 
density including boost contributions between intrinsic deuteron states
\beq
\langle d' m'|J_\lambda(0)|d\, m\rangle=2\sqrt{E_{d'}E_d}
\langle 1\, m'|\tilde J_\lambda(\vec q\,)| 1\, m\rangle\,,
\eeq
where 
\beq
\tilde J_\lambda(\vec q,\,\vec P)=\int d^3R \,U^\dagger(\vec d^{\,\prime})
J_\lambda(0)\,U(\vec d\,)e^{-i\vec q\cdot \vec R}\,,\label{tildeJ}
\eeq
with $\vec q=\vec d^{\,\prime}-\vec d$ and $\vec P=\vec d^{\,\prime}+\vec d$.
For frames collinear to $\vec q$, $\vec P$ will be proportional to $\vec q$. 
Since such frames will be considered exclusively, we can drop $\vec P$ as 
an argument in $\tilde J_\lambda$. Introducing the multipole decomposition
\beqa
\langle 1\, m'|\tilde J_\lambda(\vec q\,)| 1\, m\rangle&=&(-)^\lambda
\sqrt{2\pi(1+\delta_{\lambda 0})}\nonumber\\
&&\sum_{LM}i^L\hat L \langle 1\, m'|\Big(\delta_{\lambda 0}C_{LM}
+\delta_{|\lambda| 1}(T^{(e)}_{LM}+\lambda\,T^{(m)}_{LM})\Big)| 1\, m\rangle\,,
\eeqa
one defines the invariant charge monopole and quadrupole, and magnetic 
dipole form factors $G_C^{(c)}(Q^2)$, $G_Q^{(c)}(Q^2)$, and $G_M^{(c)}(Q^2)$, 
respectively, in terms of the reduced matrix elements 
of the corresponding operators by  
\beqa
G_C^{(c)}(Q^2)&=&\sqrt{\frac{4\pi E_{d'}^{(c)}E_d^{(c)}}{3M_d^2}}
\frac{1}{1+\eta}\,\frac{|\vec q_{lab}|}{|\vec q_{c}|}
\langle 1 \|C_0(Q^2)\|1\rangle^{(c)}\,,\label{GC_c}\\
G_Q^{(c)}(Q^2)&=&\sqrt{\frac{3\pi E_{d'}^{(c)}E_d^{(c)}}{2M_d^2}}
\frac{1}{\eta(1+\eta)}\,\frac{|\vec q_{lab}|}{|\vec q_{c}|}
\langle 1 \|C_2(Q^2)\|1\rangle^{(c)}\,,\\
G_M^{(c)}(Q^2)&=&-
\sqrt{\frac{\pi E_{d'}^{(c)}E_d^{(c)}}{M_d^2}}\,
\frac{i}{\sqrt{\eta(1+\eta)}}\, 
\langle 1 \|M_1(Q^2)\|1\rangle^{(c)}\,,
\eeqa
where we have introduced the notation $M_1(Q^2)$ instead of 
$T_1^{(m)}(Q^2)$ for the magnetic dipole operator. 
Then one finds for the longitudinal and transverse structure functions 
the well known expressions
\beqa
W_{L}^{(c)}(Q^2)&=&(1+\eta)^{2}\frac{\vec q_{c}^{\,2}}{\vec q_{lab}^{\,2}}\Big(
G_C^{(c)}(Q^2)^2+\frac{8}{9}\eta^{2}G_Q^{(c)}(Q^2)^2\Big)\,,\\
W_{T}^{(c)}(Q^2)&=&\frac{4}{3}\,\eta(1+\eta)G_M^{(c)}(Q^2)^2\,.
\eeqa
The various multipole operators are 
evaluated between intrinsic deuteron wave functions in 
the chosen frame of reference ``c''. 

With respect to the interpretation of the form factors, the following 
remark is in order. Since in the Breit frame one has 
the relation $Q^2=(\vec q_{Breit})^2$, one usually interprets 
$G_C^{(Breit)}(Q^2)$ as the Fourier transform of the charge density of the 
target. However, this interpretation is misleading, since each Fourier 
component refers to a different reference frame, because the Breit 
frame depends on the momentum transfer. Indeed, $C_0$ in (\ref{GC_c}) 
contains according to (\ref{tildeJ}) the effect of the boost of initial and 
final target states to its rest frame. In other words, the Fourier 
inversion of the form factor $G_C^{(Breit)}(Q^2)$ will not result in the 
rest frame charge density of the target, because the latter does not 
contain the boost effects. 

In principle, the superscript ``(c)'' at the form factors is redundent 
because the form factors themselves are invariant quantities, 
independent from the chosen 
reference frame if they are evaluated in a genuine covariant theory. 
However, in a nonrelativistic or semirelativistic treatment, where only 
lowest order relativistic contributions are included, they will in 
general be frame dependent, and for this reason we have kept the superscript 
``(c)'' on them. In fact, it would be quite instructive to study the frame 
dependence in such a case. For electrodisintegration this has been done 
in \cite{BeW94}. There we had introduced a so-called $\zeta$-frame allowing a
continuous variation of the reference frame between lab and antilab systems 
by variation of a parameter $\zeta$ between zero and one which corresponds
to the antilab and lab frames, respectively. The Breit 
frame is described by $\zeta=1/2$. 

In the $\zeta$-frame one has the following kinematic relations
\beq
\begin{array}{ll}
\vec d_\zeta = -(1-\zeta)\vec q_\zeta\,,&
E_{d_\zeta}=M_d\frac{1+2\eta(1-\zeta)}
{\sqrt{1+4\eta\zeta(1-\zeta)}}\,,\\
\vec d'_\zeta = \zeta\vec q_\zeta\,,& 
E_{d'_\zeta}=M_d\frac{1+2\eta\zeta}
{\sqrt{1+4\eta\zeta(1-\zeta)}}\,,
\end{array}\label{zetakin}
\eeq
with
\beq
\vec q_\zeta=\frac{\vec q_{lab}}{\sqrt{1+4\eta\zeta(1-\zeta)}}\,.
\eeq
Thus in the $\zeta$-frame the form factors become
\beqa
G_C^{(\zeta)}(Q^2)&=&\sqrt{\frac{4\pi}{3}}
\,
\frac{c_\zeta(\eta)}{1+\eta}\,
\langle 1 \|C_0(Q^2)\|1\rangle^{(\zeta)}\,,\\
G_Q^{(\zeta)}(Q^2)&=&\sqrt{\frac{3\pi}{2}}\,
\frac{c_\zeta(\eta)}{\eta(1+\eta)}\,
\langle 1 \|C_2(Q^2)\|1\rangle^{(\zeta)}\,,\\
G_M^{(\zeta)}(Q^2)&=&-i\sqrt{\frac{\pi}{\eta(1+\eta) }}\,
\frac{c_\zeta(\eta)}{\sqrt{1+4\eta\zeta(1-\zeta)}}
\langle 1 \|M_1(Q^2)\|1\rangle^{(\zeta)}\,,
\eeqa
where
\beqa
c_\zeta(\eta)&=&\sqrt{\frac{E_d E_{d'}}{M_d^2}}\,
\frac{|\vec q_{lab}|}{|\vec q_{\zeta}|}
\nonumber\\
&=&\sqrt{(1+2\eta(1-\zeta))(1+2\eta\zeta)}\nonumber\\
&=&\sqrt{1+2\eta+4\eta^2\zeta(1-\zeta)}\,
\eeqa
takes into account the noncovariant normalization and the boost from the 
$\zeta$- to the lab frame.
These expressions are symmetric with respect to the interchange 
$\zeta\leftrightarrow (1-\zeta)$ which reflects the fact, that the 
$\zeta$-frame is equivalent to the $(1-\zeta)$-frame. In particular, 
this means that lab and antilab frames are equivalent as has been 
noted before \cite{Oht88}. Thus for a check of the frame dependence 
one needs to consider only either $\zeta$ between 0 and $1/2$ or 
between $1/2$ and one. 

In the lab or antilab frames one has 
$c_0(\eta)=c_1(\eta)=\sqrt{1+2\eta}$ and thus 
\beqa
G_C^{(lab/antilab)}(Q^2)&=&\sqrt{\frac{4\pi}{3}}
\,
\frac{\sqrt{1+2\eta}}{1+\eta}\,
\langle 1 \|C_0(Q^2)\|1\rangle^{(lab/antilab)}\,,\\
G_Q^{(lab/antilab)}(Q^2)&=&\sqrt{\frac{3\pi}{2}}\,
\frac{\sqrt{1+2\eta}}{\eta(1+\eta)}\,
\langle 1 \|C_2(Q^2)\|1\rangle^{(lab/antilab)}\,,\\
G_M^{(lab/antilab)}(Q^2)&=&-i
\sqrt{\frac{\pi(1+2\eta)}{\eta(1+\eta) }}\,
\langle 1 \|M_1(Q^2)\|1\rangle^{(lab/antilab)}\,,
\eeqa
whereas in the Breit frame with $c_{\frac{1}{2}}(\eta)=1+\eta$ one finds
\beqa
G_C^{(Breit)}(Q^2)&=&\sqrt{\frac{4\pi}{3}}
\,
\langle 1 \|C_0(Q^2)\|1\rangle^{(Breit)}\,,\label{Gcbreit}\\
G_Q^{(Breit)}(Q^2)&=&\sqrt{\frac{3\pi}{2}}\,
\frac{1}{\eta}\,
\langle 1 \|C_2(Q^2)\|1\rangle^{(Breit)}\,,\\
G_M^{(Breit)}(Q^2)&=&-i\sqrt{\frac{\pi}{\eta }}\,
\langle 1 \|M_1(Q^2)\|1\rangle^{(Breit)}\,.
\eeqa
In the present work, however, we will adopt the antilab frame for the 
numerical evaluation. 

The form factors are normalized as
\beqa
G_C(0)&=&1\,,\\
G_Q(0)&=&M_d^2\,Q_d\,,\\
G_M(0)&=&\frac{M_d}{M_p}\,\mu_d\,,
\eeqa
where $\mu_d$ (in units of nuclear magnetons $\mu_N$) 
and $Q_d$ denote the static deuteron magnetic dipole and 
charge quadrupole moments, respectively, and $M_p$ the proton mass. 
Furthermore, the mean square charge radius $r_d^{ch}$ of the deuteron 
is defined by 
\beqa
(r_d^{ch})^2
&=&\left.-6\,\frac{dG_C(Q^2)}{dQ^2}\right|_{Q^2=0}
\,.
\eeqa
In the Breit frame it is usually interpreted as the mean square radius 
of the charge density (see Eq.\ (\ref{Gcbreit})). One should keep in mind 
that $r_d^{ch}$ includes the effect of finite nucleon and meson sizes. 

Instead of the representation of the differential cross section in terms
of the structure functions $W_{L/T}$ one usually introduces two other
invariant structure functions $A(Q^2)$ and $B(Q^2)$ by defining
\beqa
A(Q^2)&=&(1+\eta)^{-2}\frac{\vec q_{lab}^{\,2}}{\vec q_{c}^{\,2}}
\,W_{L}^{(c)}(Q^2)
+\frac{1}{2}(1+\eta)^{-1}W_{T}^{(c)}(Q^2)\nonumber\\
&=&G_C^{(c)}(Q^2)^2+\frac{8}{9}\eta^{2}G_Q^{(c)}(Q^2)^2+
\frac{2}{3}\,\eta\,G_M^{(c)}(Q^2)^2\,\\
B(Q^2)&=&W_{T}^{(c)}(Q^2)\nonumber\\
&=&\frac{4}{3}\,\eta(1+\eta)\,G_M^{(c)}(Q^2)^2\,,
\eeqa
and the cross section becomes
\beq
\frac{d\sigma^{lab}}{d\Omega_e^{lab}}=\sigma_{Mott}^{lab}
\Big[A(Q^2)+B(Q^2)\,\tan^2\frac{\theta_e^{lab}}{2}\,
\Big]\,.\label{diffcrossa}
\eeq
A polarization observable of considerable interest is the tensor recoil 
polarization $T_{20}$ because it is sensitive to the quadrupole form factor 
\cite{Sch64,GoP64,MoG74} according to
\beqa
T_{20}(Q^2,\theta_e^{lab})=-\frac{4\sqrt{2}\,\eta}{3S(Q^2,\theta_e^{lab})}&\,&
\Big(G_C(Q^2)\,G_Q^{(c)}(Q^2)
+\frac{\eta}{3}G_Q^{(c)}(Q^2)^2\nonumber\\
&&+
\frac{1}{8}\Big(1+2(1+\eta)\tan^2\frac{\theta_e^{lab}}{2}\Big)
\,G_M^{(c)}(Q^2)^2\Big)\,,
\eeqa
where
\beq
S(Q^2,\theta)=A(Q^2)+B(Q^2)\,\tan^2\frac{\theta}{2}\,.
\eeq
Thus the measurement of $T_{20}$ in conjunction with the structure functions 
$A(Q^2)$ and $B(Q^2)$ allows one to disentangle the charge monopole and 
quadrupole form factors.

\section{Results and discussion}
\label{results}

The various current contributions including relativistic 
terms of leading order beyond the nonrelativistic theory 
to one- and two-body charge and current density 
operators have been evaluated for the elastic form
factors, the structure functions and the tensor polarization $T_{20}$. 
We have limited our evaluation to the region of momentum transfers 
$Q^2\leq 30$ fm$^{-2}$, the reason being that in a previous study of 
electrodisintegration the limit of the nonrelativistic approach plus 
leading order relativistic contributions had been found to be roughly 
$Q^2 \sim 25$ fm$^{-2}$ \cite{BeW94}. Therefore, results for higher 
momentum transfers obtained within such a limited framework may not 
be reliable, and there any agreement with experimental data may be 
accidental and misleading. 

Our theoretical approach is based on the equation-of-motion method, 
and has been outlined in detail in \cite{GoA92}. Starting from a system of
coupled nucleon and meson fields, one eliminates the explicit meson 
degrees of freedom by the FST-method \cite{FuS54} and introduces instead 
effective operators in pure nucleonic space for both the $NN$ interaction
and the electromagnetic charge and current densities. By means of the 
Foldy-Wouthuysen transformation, one obtains the
nonrelativistic reduction including leading order relativistic
contributions. Whereas in \cite{GoA92} only pion degrees of freedom have
been considered, the extension to the three realistic Bonn OBEPQ
versions A, B, and C \cite{Mac89} 
with application to electrodisintegration has 
been reported in \cite{RiG97}. All the relevant details can be found there.
In particular, all explicit expressions are listed in the Appendix of 
\cite{RiG97} for the various one- and two-body contributions to the 
electromagnetic charge and current density operators used in this work, 
including consistently leading order relativistic terms, boost and vertex 
parts, and, furthermore, the lowest order dissociation 
$\rho\pi\gamma$-current, which is purely transverse and not fixed 
by the potential model. If not mentioned explicitly, the potental 
version OBEPQ-B is used. Each version fixes the masses, coupling strengths, 
and vertex regularization parameters for the various exchanged mesons. 
As electromagnetic nucleon form factors, we have taken the phenomenological 
dipole fit including a nonvanishing electric neutron form factor in the 
Galster parametrization \cite{GaK71}. 
For the following discussion of the effects from the various 
relativistic contributions, we use the same notational scheme
as introduced in \cite{RiG97} which we list for convenience in 
Table~\ref{short.notation}.

We will start the discussion by considering first the static 
electromagnetic properties, viz., magnetic dipole and electric 
quadrupole moments $\mu_d$ and $Q_d$, respectively, 
and the mean square charge radius $r^{ch}_d$. 
In Table~\ref{stat} the various current contributions are listed for the 
OBEPQ-B model. Relativistic one-body contributions reduce slightly 
the magnetic moment of the nonrelativistic impulse approximation (IA) by 
0.7~\%, in good agreement with \cite{TaN92}. The same magnitude but of 
opposite sign has been found in \cite{ChK89} using a covariant light-front 
approach and a $p/M$-expansion as well. The reason for this difference is 
not clear to us. Relativistic pion exchange currents including boost then
lead to an enhancement by about 1.8~\% which, however, is again weakly 
reduced by retardation ($-0.6$~\%). Further contributions of 1.7~\% from 
$\rho$-exchange, of 0.8~\% from other heavy mesons and of 0.5~\% from the 
$\rho\pi\gamma$-current result in a total increase over the 
IA by 4.2~\%. This is considerably larger than the 2.6~\% which has been 
found in \cite{WiS95} with the Argonne potential $v_{18}$. It is also 
larger than the results found in \cite{TaN92}. Comparing with the 
individual contributions of \cite{TaN92}, we find the only sizeable 
difference in the ones of $\pi$- and $\rho$-MEC, for which we obtain a total 
contribution which is larger by about a factor of two. 
The total theoretical magnetic moment of 
$0.8875\,\mu_N$ is by 3.5~\% higher than 
the experimental value of $0.8574\,\mu_N$ \cite{muexp}. 

The quadrupole moment 
is mainly affected by contributions from the pion exchange charge 
contribution giving a sizeable increase (4.9~\%) while relativistic 
one-body ($-0.7$~\%) and pion retardation ($-0.4$~\%) contributions 
yield a smaller decrease. The relativistic one-body part is in agreement
with the results of \cite{TaN92,AdG93}. With respect to the total pion 
contribution we notice a nice agreement with \cite{AdG93} but again 
our result is considerably larger than what has been found in \cite{TaN92}. 
Heavy meson exchange is almost negligible. The total effect is an 
enhancement by 3.8~\% over the IA, twice as large than found in \cite{TaN92}
and \cite{WiS95}, and the total theoretical value is in satisfactory 
agreement with the experimental value of $Q_d^{exp}= 0.2860(15)\, 
\mbox{fm}^2$ \cite{quadexp}. Finally, comparing the one-body part 
with the findings of \cite{ChK89}, one notices again the puzzling 
situation that the relativistic one-body contributions lead to 
an enhancement in the covariant approach while the $p/M$-expansion 
results in a small decrease. Whether this hints at a failure of the 
$p/M$-expansion as interpreted in \cite{ChK89} needs further studies.

The charge radius is much less sensitive to relativistic and meson exchange 
contributions due to the suppression of the short range part by the 
additional $r^2$-weighting. 
We find a total enhancement of only 0.5~\%, mainly from relativistic one-body 
and pion exchange terms, all other effects being negligible here. 
The relativistic one-body part is somewhat smaller than reported in 
\cite{BuH96} while the MEC part is of the same size than found in 
\cite{TaN92} and \cite{BuH96} but slightly larger than in \cite{Koh83}. The 
total theoretical value of 2.1121 fm is in excellent agreement with the 
earlier experimental value of 2.116(6) from \cite{SiS81} and still quite 
close to the recent experimental one of 2.127(7) fm, the difference 
being only about 1~\%. The latter value is based on an isotope shift 
experiment \cite{ScL93} resulting in 
$r_{IS}^2=(r_d^{ch})^2-(r_p^{ch})^2=3.795(19)$~fm$^2$ including first order 
nuclear structure effects. If one includes in addition second 
order nuclear structure effects from \cite{LuR93} one finds 
$r_{IS}^2=(r_d^{ch})^2-(r_p^{ch})^2=3.782(21)$~fm$^2$ as cited in a review 
by Wong \cite{Won94}. We have taken the latter value and for the proton 
$r_p^{ch}=0.862(12)$~fm from \cite{SiS81} to evaluate the experimental
deuteron charge radius listed in Tables~\ref{stat} and \ref{statabc}. 

The potential model dependence of the static e.m.\ deuteron properties
is shown in Table~\ref{statabc}. The largest relative variation is 
found for the quadrupole moment of about 1.2~\% increase going from 
model A to B and to C which is correlated with the increase of the 
tensor force as is indicated by the 
$D$-state percentage $P_d$ listed also in Table~\ref{statabc}. 
Similarly, one finds a steady increase of the magnetic moment with 
increasing $P_d$, each time of about 0.5~\% going from model A and B to C. 
Only the charge radius is almost independent from the potential model. 

\begin{figure}
\centerline{%
\epsfxsize=13.0cm
\epsffile{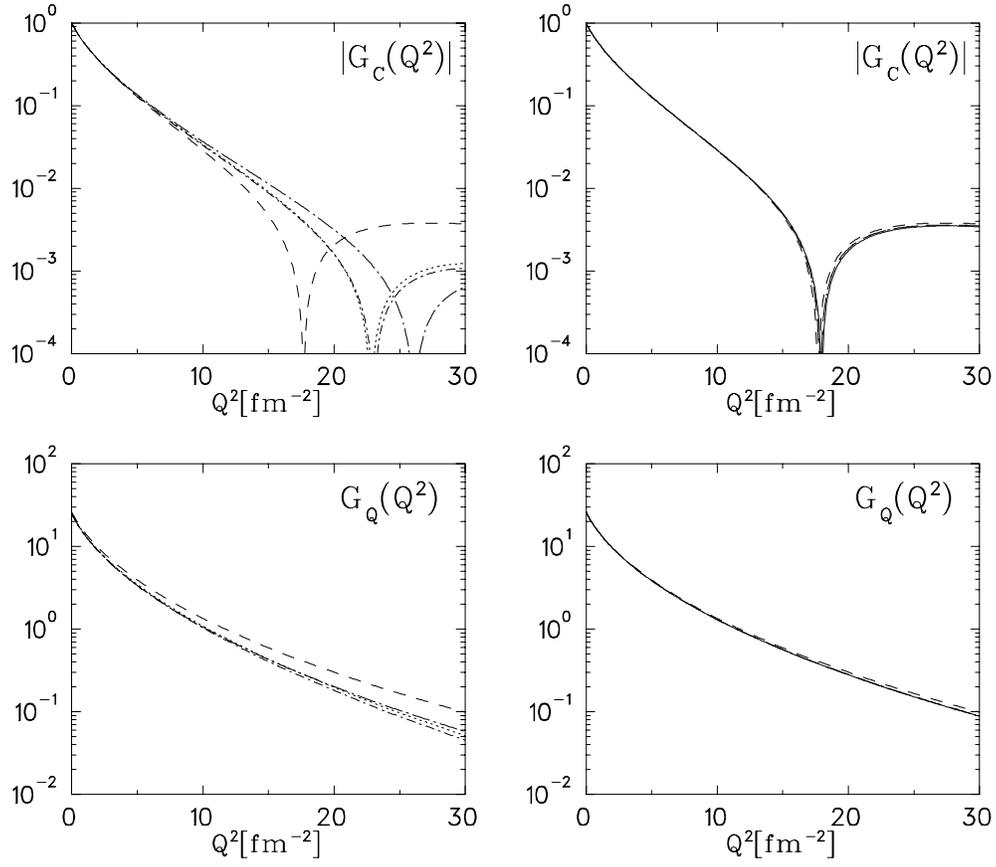}
}
\vspace*{1cm}
\caption{Various current contributions to charge monopole and 
quadrupole form factors $G_C(Q^2)$ and $G_Q(Q^2)$, respectively, 
for the Bonn OBEPQ-B potential model 
as function of $Q^2$. Notation of the 
curves in the left panels: 
dotted: $n$; short-dash-dot: $n(r)$; long-dash-dot: $n(r,\chi_0)$; 
short dashed: $n(r,\chi_0)\pi(r)$; 
in the right panels: 
short dashed: $n(r,\chi_0)\pi(r)$; long dashed: $n(r,\chi_0)\pi(r,t)$; 
wide dotted: $n(r,\chi_0)\pi(r,t,\chi_0,\chi_V)$; solid: total.  
\label{GCQ}}
\end{figure}

\begin{figure}
\centerline{%
\epsfxsize=13.0cm
\epsffile{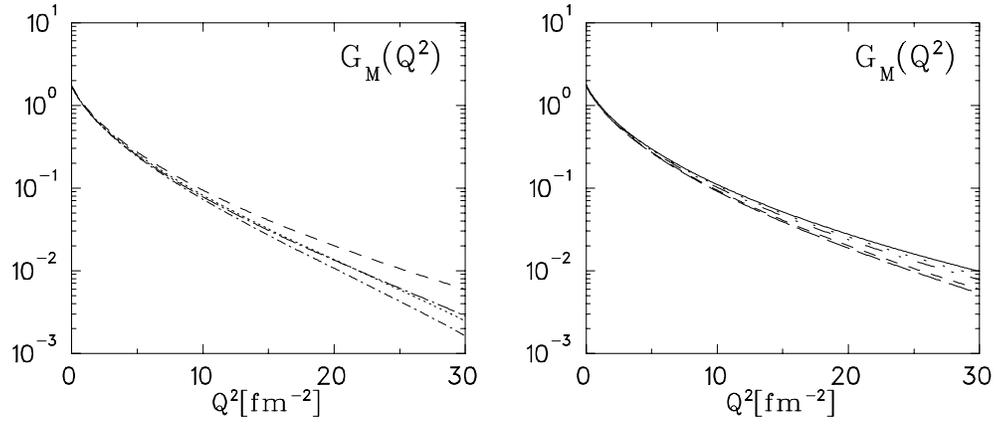}
}
\vspace*{1cm}
\caption{
Various current contributions to magnetic dipole form factor 
$G_M(Q^2)$ 
for the Bonn OBEPQ-B potential model as function of $Q^2$. Notation of the 
curves in the left panel: 
dotted: $n$; 
short-dash-dot: $n(r)$; 
long-dash-dot: $n(r,\chi_0)$; 
short dashed: $n(r,\chi_0)\pi(r)$; 
in the right panel: 
short dashed: $n(r,\chi_0)\pi(r)$; 
long dashed: $n(r,\chi_0)\pi(r,t,\chi_0,\chi_V)$; 
wide dash-dot: $n(r,\chi_0)\pi(r,t,\chi_0,\chi_V)\rho$;
wide dotted: $n(r,\chi_0)\pi(r,t,\chi_0,\chi_V)\rho(\chi_0)h(\chi_0)$; 
solid: total.
\label{GM}}
\end{figure}

As next we will discuss the results for the form factors. The influence 
of various contributions are shown in Fig.~\ref{GCQ} for the charge 
and quadrupole form factors and in Fig.~\ref{GM} for the magnetic 
dipole form factor. In the left panels we show separately the effects of
the relativistic one-body currents, of the one-body boost, and of the 
relativistic $\pi$-MEC. The additional effects from retardation and 
two-body boost for the $\pi$-MEC, from heavy meson exchange and from the 
$\rho\pi\gamma$-current are exhibited separately in the right panels. 
Looking first at the upper left panel of Fig.~\ref{GCQ}, one readily 
notices that the relativistic 
one-body contributions to the charge density show only a very tiny effect 
on $G_C(Q^2)$ while the corresponding one-body boost results 
in quite an enhancement and a sizeable shift of the zero towards 
higher momentum transfer. The significance of the one-body boost has 
already been observed before in \cite{TaN92,AdG93} and and our results 
are in qualitative agreement with those findings. We would like to 
emphasize that neglect of this boost part as in the calculations of 
\cite{ScR91,WiS95} leads to a significant overestimation. 

Adding now the relativistic $\pi$-MEC, yields a strong reduction. In 
fact, it is by far the largest effect which reverses the upshift of the zero 
by relativistic one-body contributions and shifts it down beyond 
the nonrelativistic one. This important effect of the $\pi$-MEC is in 
accordance with earlier findings in, e.g., \cite{MoR89,ScR91,WiS95,HaM80} 
and in particular with \cite{HeA95} and \cite{PlC95}. 
All other additional contributions, as shown in the upper right panel 
of Fig.~\ref{GCQ}, from $\pi$-retardation, two-body boost, heavy meson 
exchange are very small in this region of $Q^2$. The 
$\rho\pi\gamma$-current does not contribute, because we have considered 
its lowest order contribution only which is purely transverse. 
The situation for $G_Q(Q^2)$ in the 
lower left panel of Fig.~\ref{GCQ} is different 
with respect to the relative importance of the various contributions. 
Here all relativistic one-body terms show much less effects compared to 
$G_C(Q^2)$, only $\pi$-MEC gives a sizeable increase of the form factor 
in accordance with the results in \cite{HeA95}. 
Again all other further contributions have an almost negligible influence on 
$G_Q$, as is seen in the lower right panel of Fig.~\ref{GCQ}. 

For the magnetic dipole form factor $G_M(Q^2)$ the role of the various 
contributions is markably different as can be seen in Fig.~\ref{GM}. 
In the left panel one notices that already the relativistic one-body 
currents lead to a significant decrease of the nonrelativistic IA. However, 
the one-body boost contributions are sizeable too, 
but act in the opposite direction so that the total result is very close 
to the original IA. Again the remark is in order that leaving out the 
one-body boost contributions leads to an underestimation. The inclusion 
of the relativistic $\pi$-MEC yields then a drastic 
increase with increasing momentum transfer. Adding retardation and boost for 
$\pi$-exchange, shown in the right panel, results in a slight reduction 
which is more than compensated by $\rho$-MEC. Further slight increases 
come from the additional contributions of other heavy meson exchanges 
and from the dissociation $\rho\pi\gamma$-current. 

\begin{figure}
\centerline{%
\epsfxsize=13.0cm
\epsffile{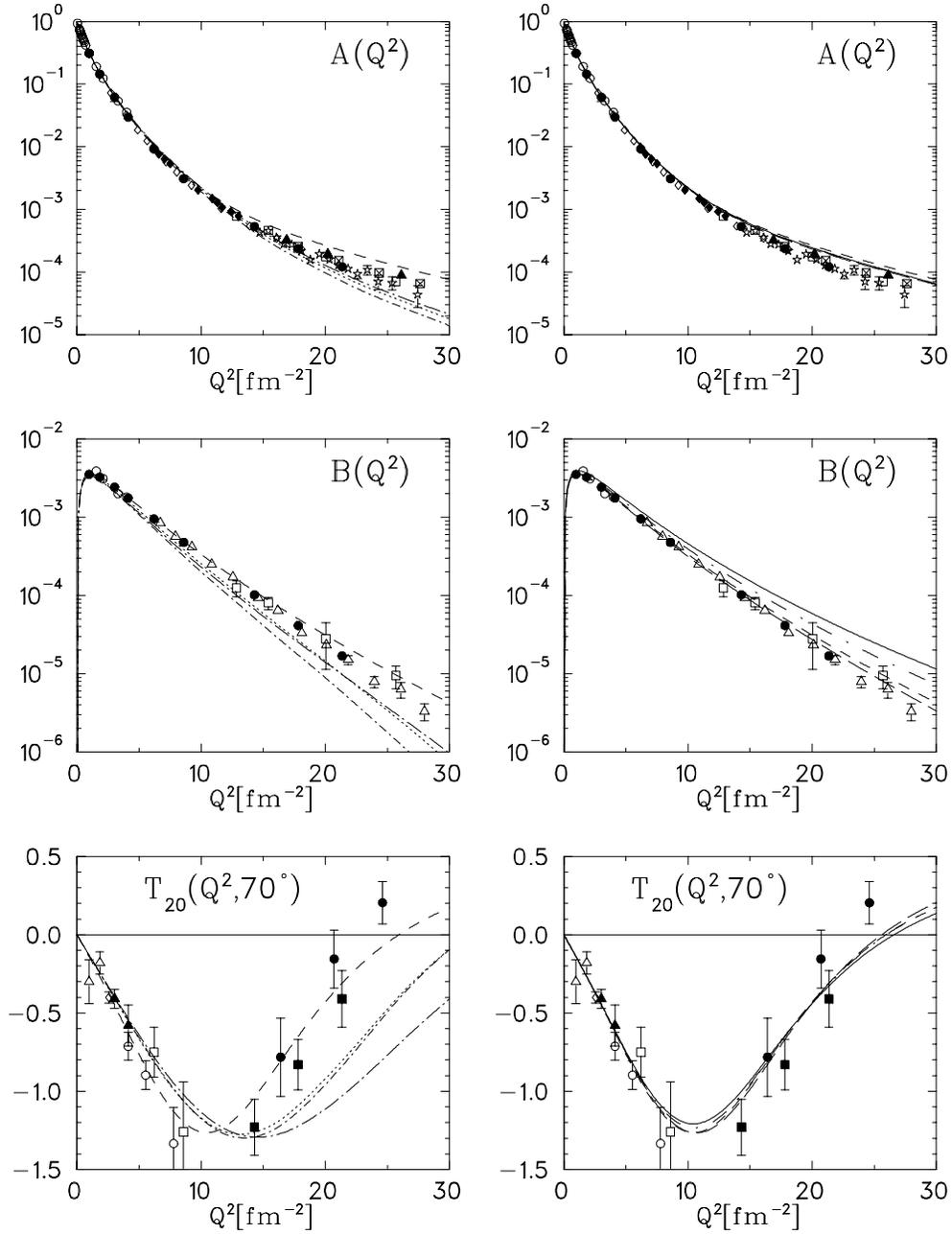}
}
\vspace*{1cm}
\caption{
Various current contributions to the structure functions $A(Q^2)$, 
$B(Q^2)$, and the tensor polarization $T_{20}(Q^2,70^\circ)$
for the Bonn OBEPQ-B potential model as function of $Q^2$. 
Notation of the curves in the left panels: 
dotted: $n$; short-dash-dot: $n(r)$; long-dash-dot: $n(r,\chi_0)$; 
short dashed: $n(r,\chi_0)\pi(r)$; 
in the right panels: 
short dashed: $n(r,\chi_0)\pi(r)$; long dashed: $n(r,\chi_0)\pi(r,t)$; 
wide dotted: $n(r,\chi_0)\pi(r,t,\chi_0,\chi_V)$; solid: total.  
Experimental data for $A(Q^2)$ and $B(Q^2)$: 
open stars: \protect\cite{ElF69};
full diamonds: \protect\cite{GaK71};
open circles: \protect\cite{SiS81}; 
open squares: \protect\cite{Cra85};
open triangles: \protect\cite{AuC85};
open diamonds: \protect\cite{PlA90};
full circles \protect\cite{Gar94}; 
full triangles: \protect\cite{Abb99};
Experimental: data for $T_{20}(Q^2,70^\circ)$: 
full triangles: \protect\cite{Sch84};
open triangles: \protect\cite{Dmi85}; 
open squares: \protect\cite{Gil90}; 
full squares: \protect\cite{The91}; 
open diamond: \protect\cite{FeB96};
full circles: \protect\cite{Fur98} (preliminary); 
open circles: \protect\cite{BoA99}. 
\label{ABT}
}
\end{figure}

The observables, i.e., the structure functions $A(Q^2)$ and $B(Q^2)$, 
and the tensor polarization $T_{20}(Q^2,70^\circ)$ are displayed in 
Fig.~\ref{ABT}, again with separate contributions shown in the left and 
right panels as for the form factors in the previous figures. 
For $A(Q^2)$, shown in the upper left panel, the relativistic one-body 
current gives a lowering of the IA above $Q^2\sim 10$ fm$^{-2}$, which 
is counteracted by the corresponding boost terms so that the net result is 
only a tiny enhancement of the IA. A stronger increase is generated by the 
relativistic $\pi$-MEC, again above $Q^2\sim 10$ fm$^{-2}$. 
All other additional contributions from retardation, two-body boost, 
heavy meson exchange and 
$\rho\pi\gamma$-current, shown in the upper right panel, are very small. 
Compared to the experimental data, also displayed in Fig.~\ref{ABT}, 
one notices a satisfactory agreement for momentum transfers 
$Q^2\leq 10$ fm$^{-2}$. For higher $Q^2$ the IA including relativistic 
one-body contributions lies systematically below the data, while 
addition of relativistic $\pi$-MEC results in a systematic overshooting 
which is not compensated enough by the slight reduction from the heavy meson 
and $\rho\pi\gamma$-currents. This overshooting has also been observed in 
\cite{PlC95}. 

The behaviour of $B(Q^2)$, shown in the middle left and right panels, 
reflects the one of $G_M(Q^2)$ in Fig.~\ref{GM} and thus we do not need 
to discuss the various relativistic contributions again in detail. 
With respect to experiment, already at rather low momentum transfers, 
the IA plus relativistic 
one-body contributions lies below the experimental data. Here the inclusion 
of the relativistic $\pi$-MEC leads to quite a satisfactory agreement 
although the theory lies systematically slightly higher than the data. 
It corresponds to what has been noted above for the static magnetic moment
$\mu_d$. However, the further contributions from the heavy meson and 
$\rho\pi\gamma$-currents spoil this nice agreement and lead to a 
sizeable overestimation, considerably larger than for $A(Q^2)$. 
Such a systematic overprediction has also been 
observed by Plessas {\it et al.}~\cite{PlC95} as well as by Wiringa et 
al.~\cite{WiS95} although in the latter case of somewhat smaller size, but 
they had left out the boost contributions which would have shifted further 
up their results. The origin of this systematic disagreement is an open
question and it needs further detailed studies to clarify it. 
Obviously, it appears that the magnetic properties of 
hadronic systems are more sensitive to finer details of the dynamic 
interaction effects than charge ones, for which charge conservation 
constitutes a stringent condition. It is interesting to note that the 
recent relativistic calculation of Phillips {\it et al.}~\cite{PhW99} results 
in a systematic underestimation of the data for both structure functions 
in this region of momentum transfers. 

The two lower panels of Fig.~\ref{ABT} show the influence of the various 
current contributions on the tensor polarization $T_{20}(Q^2,70^\circ)$. 
Since the 
minimum is determined by the zero of $G_C(Q^2)$, one readily notices that the 
shift of the minimum by the different current contributions follows the 
shift of the zero of $G_C(Q^2)$, i.e., the largest effects come from the 
one-body boost shifting the minimum to higher $Q^2$ and from the 
$\pi$-exchange resulting in an even larger downshift of the minimum and 
of the zero crossing of $T_{20}(Q^2,70^\circ)$. With respect to the 
experimental data, 
the one-body currents alone yield a drastic disagreement for momentum 
transfers above the minimum, whereas inclusion of the relativistic $\pi$-MEC
gives an almost satisfactory description although the data seem to rise 
slightly steeper at higher momentum transfers. All other further contributions 
show almost no influence at all. 

\begin{figure}
\centerline{%
\epsfxsize=13.0cm
\epsffile{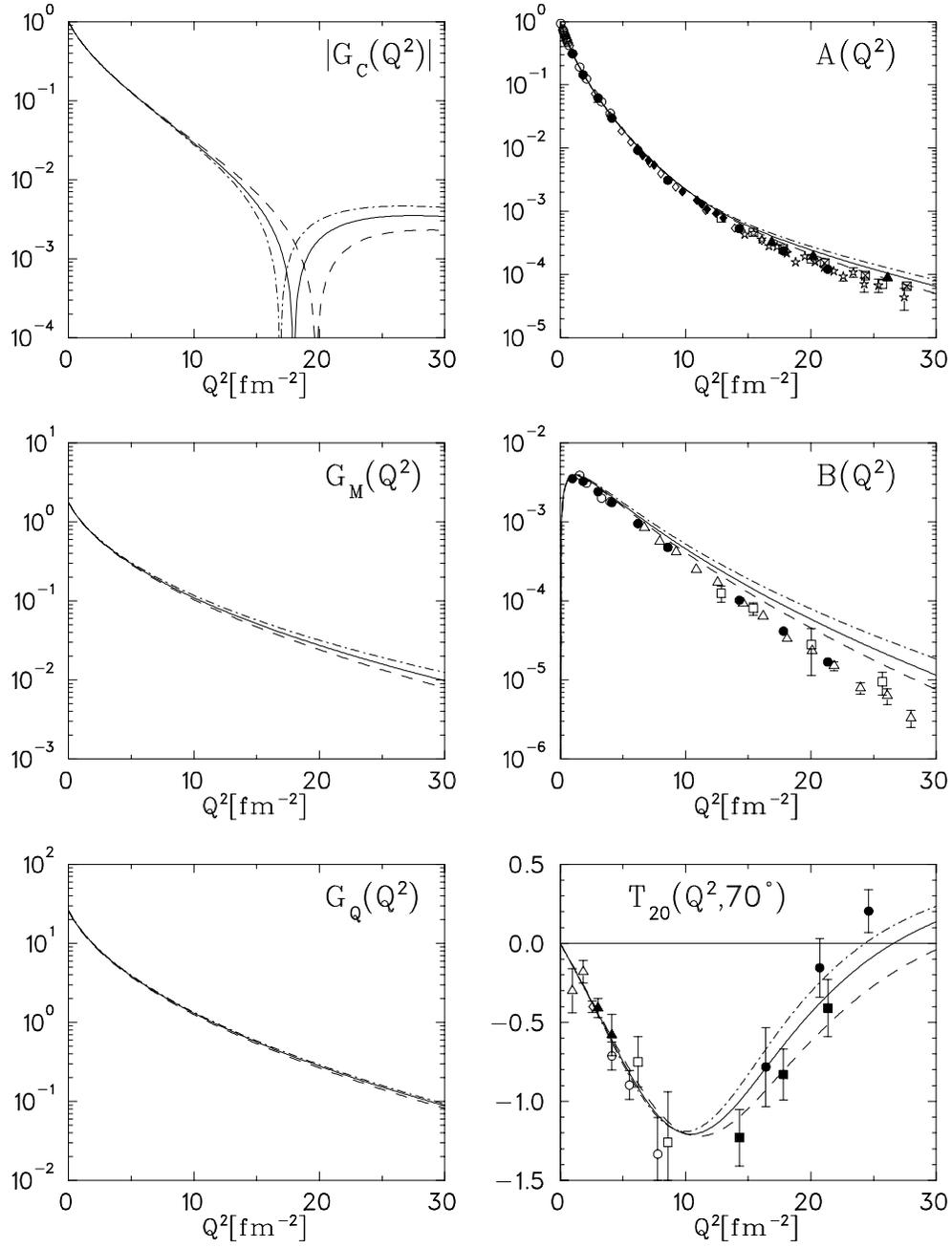}
}
\vspace*{1cm}
\caption{Comparison of the three Bonn OBEPQ potential versions for the 
form factors and observables for the total current contributions. 
Notation of the curves: 
dashed: OBEPQ-A; 
solid: OBEPQ-B;
dash-dot: OBEPQ-C. 
Experimental data as in Fig.~\ref{ABT}.
\label{GCMQABC}
}
\end{figure}

Finally we show in Fig.~\ref{GCMQABC} the dependence of form factors and 
observables on the three Bonn OBEPQ versions A, B, and C. The largest 
dependence on the potential model is seen in $G_C(Q^2)$ above 
$Q^2\sim 10$ fm$^{-2}$ via the shift of the zero. With increasing 
$D$-state probability $P_d$ the zero is shifted downward. In contrast 
to this, $G_Q(Q^2)$ is rather insensitive to the potential model. This 
different behaviour is also reflected in the observables $A(Q^2)$ and 
$T_{20}(Q^2,70^\circ)$. The structure function $A(Q^2)$ in the upper 
right panel is almost model independent for $Q^2\leq 10$ fm$^{-2}$ while 
above this region some model dependence is seen which increases with 
increasing $Q^2$. Also the tensor polarization $T_{20}(Q^2,70^\circ)$ 
in the lower right panel exhibits very little model dependence up to 
$Q^2\sim 10$ fm$^{-2}$, but at higher momentum transfers one notes a 
sizeable shift of the zero and a corresponding shift of the minimum 
towards higher $Q^2$ with increasing strength of the tensor force, i.e., 
increasing $P_d$, of the potential model. The magnetic form factor 
$G_M(Q^2)$ and thus $B(Q^2)$, shown in the middle left and 
right panels, respectively, exhibit an even larger sensitivity to the 
potential model over the whole region of momentum transfers which 
increases in size. The absolute values are correlated with the strength 
of the tensor force, i.e., one finds the highest values for the potential 
model C having the strongest tensor force. 

\begin{figure}
\centerline{%
\epsfxsize=12.0cm
\epsffile{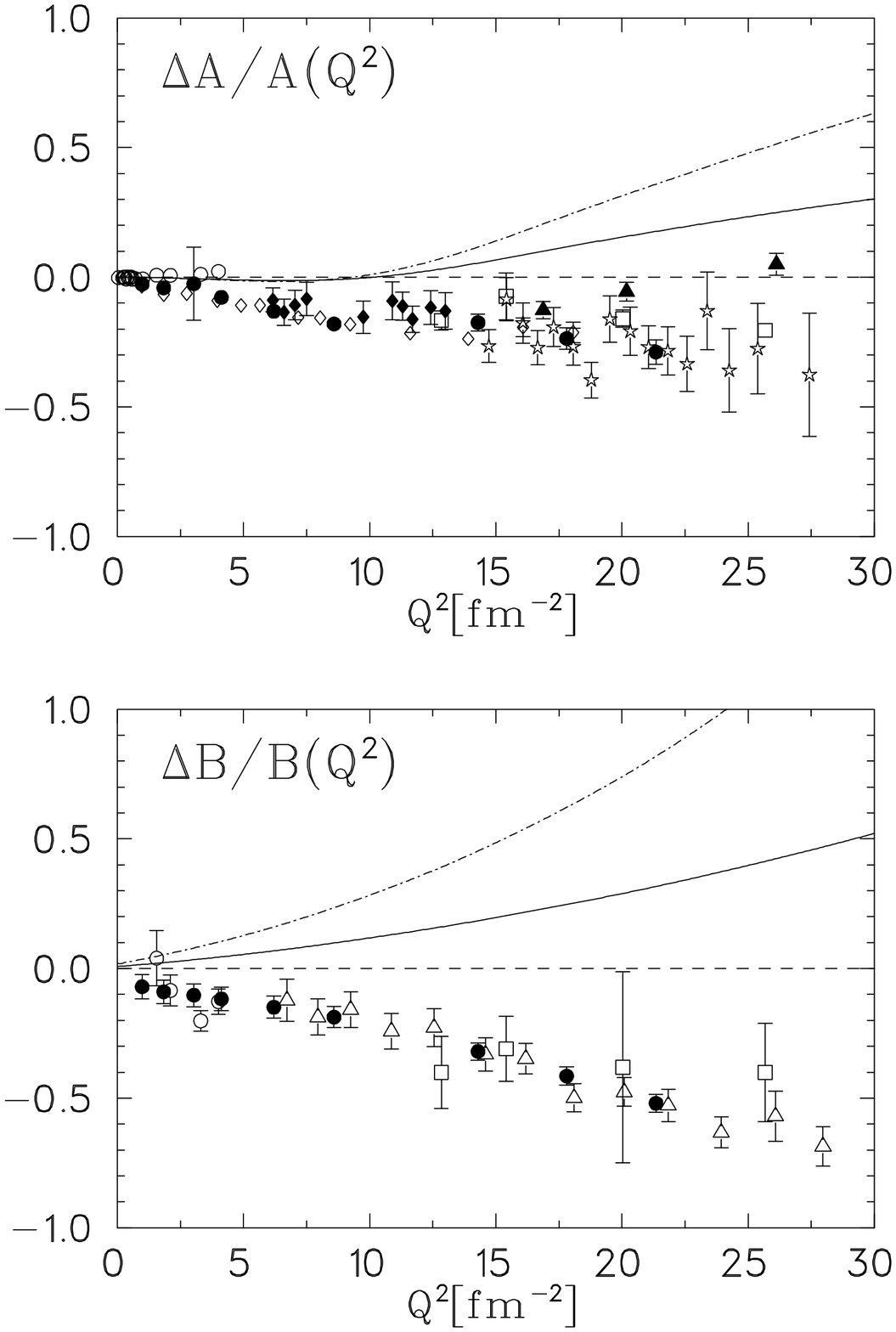}
}
\vspace*{1cm}
\caption{Relative deviation of the structure functions $A(Q^2)$ and 
$B(Q^2)$ from the theoretical prediction for the Bonn OBEPQ-A potential 
including all current contributions. 
Notation of the curves: 
dashed: OBEPQ-A; 
solid: OBEPQ-B;
dash-dot: OBEPQ-C. 
Experimental data as in Fig.~\ref{ABT}.
\label{RelAB}
}
\end{figure}

We furthermore show in Fig.~\ref{GCMQABC} a comparison with the 
experimental data for the observables. In order to exhibit more clearly the 
deviations between the three potential versions and from the experimental 
data we show in addition separately for the structure functions 
$A(Q^2)$ and $B(Q^2)$ in Fig.~\ref{RelAB} the relative deviation with 
respect to the theoretical predictions of the OBEPQ-A model including all 
current contributions. As already stated, at low momentum transfers 
($Q^2\le 10$ fm$^{-2}$) the potential model dependence of $A(Q^2)$ is 
negligible, and all potential models give a satisfactory agreement 
with the data. Above $Q^2\sim 10$ fm$^{-2}$, however, increasing 
overestimation of the data is seen for $A(Q^2)$ for all three models, 
the largest deviation being for the OBEPQ-C model having the largest $P_d$, 
whereas for the OBEPQ-A model with the smallest $P_d$ the theory is much  
closer to the data. For $B(Q^2)$ one notices already at low momentum 
transfers a systematic overestimation of the data by all three models in 
accordance with the overestimation of the static magnetic dipole moment. 
The deviation increases noticably with increasing momentum transfers. 
Again one readily sees a strong correlation between the size of the 
deviation and the size of $P_d$. As has been remarked already above with 
respect to the results with the potential model B, one finds for all 
three potential models a considerably larger overprediction for $B(Q^2)$ 
compared to $A(Q^2)$. 

Within the experimental errors $T_{20}(Q^2,70^\circ)$ is well described 
by the predictions for the OBEPQ-B model, but also the results for the 
other two OBEPQ models are in reasonable agreement in view of the large 
error bars. The new data above 15 fm$^{-2}$ from \cite{Fur98} seem to 
favour the OBEPQ-C model but one has to await the final data analysis
before one could make such a definite conclusion. Considering all data, 
the best overall description is achieved with the OBEPQ-B model at present. 
But, as already mentioned, above $Q^2\sim 15$~fm$^2$ the data seem to rise 
steeper. Thus a more precise location of the zero crossing would be most 
interesting. 

In conclusion we may state that the consistent inclusion of all 
important relativistic contributions of leading order within a 
pure nucleonic one-boson-exchange model leads to a satisfactory description 
of the structure functions of elastic electron-deuteron scattering at low 
and medium momentum transfers ($Q^2\leq 30$~fm$^2$) for the OBEPQ-A model, 
if one satisfies oneself with an agreement between theory and experiment 
within the level of about ten to twenty percent for $A(Q^2)$, whereas for 
$B(Q^2)$ the deviation grows considerably larger, e.g., a factor two at 
$Q^2 = 25$~fm$^2$. The tensor polarization 
is better described with the OBEPQ-B model. If, however, one aimes at a 
more precise agreement at the level of one or even less than one percent, 
significant further improvements are needed, in particular a much better 
description for the magnetic properties will be a very challenging task 
for the future. One might speculate whether explicit isobar 
\cite{FaA74,WeA78,ObP92} 
or even quark-gluon degrees of freedom \cite{BuY89} might lead to a 
significant improvement or whether other off-shell effects in 
electromagnetic form factors of the various currents may be important.

We would like to add another remark concerning the quality of the Bonn 
OBEPQ potentials, considered in this work, compared to more recent 
so-called ``high quality'' $NN$-potentials. These latter potentials 
use a slightly smaller $\pi N$-coupling constant than the one taken in 
the above OBEPQ potentials \cite{Mac99}. This affects in particular the 
strength of the tensor force becoming somewhat weaker in these modern 
$NN$-potentials which in turn would lead to a smaller quadrupole moment 
of the deuteron. A way out of this dilemma is discussed in \cite{Mac99} 
by assuming a charge splitting of the $\pi N$-coupling constant. Therefore, 
it remains as a task for the future to study these high quality potentials
in elastic electron deuteron scattering. Due to the phenomenological
character of these potentials, the construction of a consistent current is
quite involved. Our results, however, underline the necessity of such a 
consistent calculation, since otherwise any conclusions drawn would be rather
premature.

\begin{table}
\caption{%
\label{short.notation}
Explanation of the notation used in the tables and figure captions.}
\begin{tabular}{ll}
notation & explanation \\
\hline\hline
$n$ & nonrelativistic nucleon current\\
$n(r)$ & relativistic nucleon current \\
$n(r,\chi_0)$ & relativistic nucleon current including 
kinematic boost currents \\
\hline
$\pi(r)$ & static relativistic $\pi$-MEC \\
$\pi(r,t)$ & $\pi(r)$ + retardation contributions \\
$\pi(r,t,\chi_0,\chi_V)$ & $\pi(r,t)$
 + kinematic and potential dependent boost currents \\
\hline
$\rho(\chi_0)$ & full $\rho$-MEC  + kinematic boost currents \\
\hline
$h(\chi_0)$ & heavy meson exchange currents
 ($\eta, \omega, \sigma, \delta$) + kinematic boost currents \\
$d$    &  $\rho\pi\gamma$-current \\
\hline
total & $n(r,\chi_0)\pi(r,t,\chi_0,\chi_V)
\rho(\chi_0)h(\chi_0)d$
\end{tabular}
\end{table}

\begin{table}
\caption{%
Static properties of the deuteron (magnetic dipole moment $\mu_d$, electric 
quadrupole moment $Q_d$, and mean square charge radius $r_d^{ch}$) 
in various approximations for the Bonn OBEPQ-B model. 
\label{stat}}
\begin{tabular}{cccc}
ingredient & $\mu_d\,[\mu_N]$ & $Q_d\,[\mbox{fm}^2]$ & $r_d^{ch}\,[\mbox{fm}]$\\
\hline
 $n$                                & 0.8515   & 0.2780   & 2.1016   \\
 $n(r,\chi_0)$                      & 0.8457   & 0.2762   & 2.1073   \\ 
$n(r,\chi_0)\pi(r)$                 & 0.8673   & 0.2899   & 2.1120   \\ 
$n(r,\chi_0)\pi(r,t,\chi_0,\chi_V)$ & 0.8624   & 0.2888   & 2.1122   \\ 
$n(r,\chi_0)\pi(r,t,\chi_0,\chi_V)\rho(\chi_0)$ 
                                    & 0.8767   & 0.2888   & 2.1122   \\ 
$n(r,\chi_0)\pi(r,t,\chi_0,\chi_V)\rho(\chi_0)h(\chi_0)$
                                    & 0.8833   & 0.2886   & 2.1121   \\ 
total                               & 0.8875   & 0.2886   & 2.1121   \\ 
\hline
experiment & 0.857438230(24) \cite{muexp} & 0.2860(15) \cite{quadexp} &  
2.116(6) \cite{SiS81}\\
 & & & 2.127(7)$^{\rm a}$\\
\end{tabular}
{\small $^{\rm a}$from isotope shift experiment, for the reference see text.}
\end{table}

\begin{table}
\caption{%
Static properties of the deuteron (magnetic dipole moment $\mu_d$, electric 
quadrupole moment $Q_d$, mean square charge radius $r_d^{ch}$) 
including all leading order relativistic contributions, and the 
$D$-wave percentage $P_d$ for the various Bonn OBEPQ models. 
\label{statabc}}
\begin{tabular}{ccccc}
potential model & $\mu_d\,[\mu_N]$ & $Q_d\,[\mbox{fm}^2]$ & 
$r_d^{ch}\,[\mbox{fm}]$ & $P_d$ [\%]\\
\hline
OBEPQ-A                        & 0.8841   & 0.2850   & 2.1120 & 4.38  \\ 
OBEPQ-B                        & 0.8875   & 0.2886   & 2.1121 & 4.99  \\ 
OBEPQ-C                        & 0.8917   & 0.2917   & 2.1112 & 5.61  \\ 
\hline
experiment & 0.857438230(24) \cite{muexp} & 0.2860(15) \cite{quadexp} &  
2.116(6) \cite{SiS81} &\\
 & & & 2.127(7)$^{\rm a}$\\
\end{tabular}
{\small $^{\rm a}$from isotope shift experiment, for the reference see text.}
\end{table}

\newpage

\begin{references}

\bibitem{RiG97} 
F.\ Ritz, H.\ G\"oller, T.\ Wilbois, and H.\ Arenh\"ovel,
Phys.\ Rev.\ C {\bf 55}, 2214 (1997).

\bibitem{RiA98}
F.\ Ritz, H.\ Arenh\"ovel, and T.\ Wilbois, 
Few-Body Syst.\ {\bf 24}, 123 (1998).

\bibitem{MHE87}
R.\ Machleidt, K.\ Holinde, and Ch.\ Elster, 
Phys.\ Rep.\ {\bf 149}, 1 (1987).

\bibitem{Mac89}
R.\ Machleidt, Adv.\ Nucl.\ Phys.\ {\bf 19}, 189 (1989). 

\bibitem{Gro91}
F. Gross, in {\it Modern Topics in Electron Scattering},
eds. B. Frois and I. Sick (World Scientific, Singapore, 1991) p.\ 219.

\bibitem{Wal98}
S.J. Wallace, Nucl.\ Phys.\ A {\bf 631}, 137c (1998). 

\bibitem{PhW99}
D.R. Phillips, S.J. Wallace, and N.K. Devine, nucl-th/9906086. 

\bibitem{Gou66}
M. Gourdin, {\it Diffusion des \'Electrons de Haute \'Energie} 
(Masson, Paris, 1966).

\bibitem{Lev74}
J.S. Levinger, Springer Tracts in Modern Physics, Vol. {\bf 71}, p. 88 
(Springer-Verlag, Berlin, 1974).

\bibitem{Cio80}
C. Ciofi degli Atti, Prog. Part. Nucl. Phys. {\bf 3}, 163 (1980).

\bibitem{MoR89}
B.\ Mosconi and P.\ Ricci, Few-Body Syst.\ {\bf 6}, 63 (1989).

\bibitem{BuY89}
A.J. Buchmann, Y. Yamauchi, and A. Faessler, 
Nucl.\ Phys.\ A {\bf 496}, 621 (1989). 

\bibitem{ScR91}
R.\ Schiavilla and D.O.\ Riska,
Phys.\ Rev.\ C {\bf 43}, 437 (1991). 

\bibitem{TaN92}
K.\ Tamura, T.\ Niwa, T.\ Sato, and H.\ Ohtsubo,
Nucl.\ Phys.\ {\bf A536}, 597 (1992).

\bibitem{AdG93}
J. Adam, Jr., H. G\"oller, and H. Arenh\"ovel, Phys.\ Rev.\ C {\bf 48}, 
370 (1993).

\bibitem{WiS95}
R.B.\ Wiringa, V.G.\ Stoks, and R.\ Schiavilla, 
Phys.\ Rev.\ C {\bf 51}, 38 (1995).

\bibitem{HeA95}
H. Henning, J. Adam, Jr., P.U. Sauer, and A. Stadler, 
Phys.\ Rev.\ C {\bf 52}, R471 (1995). 

\bibitem{PlC95}
W. Plessas, V. Christian, and R.F. Wagenbrunn, 
Few-Body Syst. Suppl. {\bf 9}, 429 (1995).

\bibitem{BeW94}
G.\ Beck, T.\ Wilbois, and H.\ Arenh\"ovel,
Few-Body Syst.\ {\bf 17}, 91 (1994).

\bibitem{Oht88}
K.\ Ohta, J.\ Phys.\ G {\bf 14}, 449 (1988).

\bibitem{Sch64}
D. Schildknecht, Phys. Lett. {\bf 10}, 254 (1964); 
Z. Phys. {\bf 185}, 382 (1965).

\bibitem{GoP64}
M. Gourdin and C.A. Piketty, Nuovo Cimento {\bf 32}, 1137 (1964).

\bibitem{MoG74}
M.J. Moravcsik and P. Ghosh, Phys.\ Rev.\ Lett.\ {\bf 32}, 321 (1974).

\bibitem{GoA92}
H.\ G\"oller and H.\ Arenh\"ovel,
Few-Body Syst.\ {\bf 13}, 117 (1992).

\bibitem{FuS54}
N.\ Fukuda, K.\ Sawada, and M.\ Taketani, Progr.\ Theor.\ Phys.\ 
{\bf 12}, 156 (1954).

\bibitem{AT89}
J.\ Adam, Jr., E.\ Truhlik, and D.\ Adamova,
 Nucl.\ Phys.\ {\bf A492}, 556 (1989).

\bibitem{GaK71}
S.\ Galster, H.\ Klein, J.\ Moritz, K.H.\ Schmidt, D.\ Wegener,
and J.\ Bleckwenn,  Nucl.\ Phys.\ {\bf B32}, 221 (1971).

\bibitem{muexp}
E.R. Cohen and B.N. Taylor, Rev. Mod. Phys. {\bf 59}, 1121 (1987).

\bibitem{quadexp}
D.M. Bishop and L.M. Cheung, Phys.\ Rev.\ A {\bf 20}, 381 (1979);
T.E.O. Ericson and M. Rosa-Clot, Nucl.\ Phys.\ {\bf 405}, 497 (1983).

\bibitem{SiS81}
G.G.\ Simon, Ch.\ Schmitt, and V.H.\ Walther, 
Nucl.\ Phys.\ A {\bf 364}, 285 (1981).

\bibitem{ChK89}
P.L. Chung, B.D. Keister, and F. Coester, 
Phys.\ Rev.\ C {\bf 39}, 1544 (1989).

\bibitem{BuH96}
A.J. Buchmann, H. Henning, and P.U. Sauer, 
Few-Body Syst.\ {\bf 21}, 149 (1996).

\bibitem{Koh83}
M. Kohno, J. Phys. G {\bf 9}, L85 (1983).  

\bibitem{ScL93}
F.\ Schmidt-Kaler, D.\ Leibfried, M.\ Weitz, and T.W.\ H\"ansch, 
Phys.\ Rev.\ Lett.\ {\bf 70}, 2261 (1993).

\bibitem{LuR93}
Y. Lu and R. Rosenfelder, Phys. Lett. B {\bf 319}, 7 (1993);  
Phys. Lett. B {\bf 333}, 564 (1994) (E). 

\bibitem{Won94}
C.W. Wong, Int. J. Mod. Phys. E {\bf 3}, 821 (1994).

\bibitem{HaM80}
M.I.\ Haftel, L.\ Mathelitsch, and H.F.K.\ Zingl, 
Phys.\ Rev.\ C {\bf 22}, 1285 (1980).

\bibitem{ElF69}
J.E. Elias, J.L. Friedman, G.C. Hartmann, H.W. Kendall, P.N. Kirk, 
M.R. Sogard, L.P. Van Speybroeck, and J.K. de Pagter, 
Phys. Rev. {\bf 177}, 2075 (1969).

\bibitem{Cra85}
R. Cramer {\it et al.}, Z.\ Phys.\ C {\bf 29}, 513 (1985).

\bibitem{AuC85}
S. Auffret {\it et al.}, Phys. Rev. Lett. {\bf 54}, 649 (1985).

\bibitem{PlA90}
S. Platchkov {\it et al.}, Nucl. Phys. A {\bf 510}, 740 (1990).


\bibitem{Gar94}
M.\ Gar\c{c}on {\it et al.}, Phys.\ Rev.\ C {\bf 49},  2516 (1994).  

\bibitem{Abb99}
D. Abbott {\it et al.}, Phys. Rev. Lett. {\bf 82}, 1379 (1999).

\bibitem{Sch84}
M.E.\ Schulze {\it et al.}, Phys.\ Rev.\ Lett.\ {\bf 52}, 597 (1984).

\bibitem{Dmi85}
V.F.\ Dmitriev {\it et al.}, Phys.\ Lett.\ {\bf 157B}, 143 (1985).

\bibitem{Gil90}
R.\ Gilman {\it et al.}, Phys.\ Rev.\ Lett.\ {\bf 65}, 1733 (1990).

\bibitem{The91}
I.\ The {\it et al.}, Phys.\ Rev.\ Lett.\ {\bf 67}, 173 (1991).

\bibitem{FeB96}
M. Ferro-Luzzi {\it et al.}, Phys. Rev. Lett. {\bf 77}, 2630 (1996).

\bibitem{Fur98}
C. Furget {\it et al.}, Acta Phys. Pol. B {\bf 29}, 3301 (1998). 

\bibitem{BoA99}
M. Bouwhuis {\it et al.}, Phys. Rev. Lett. {\bf 82}, 3755 (1999).

\bibitem{FaA74}
W. Fabian, H. Arenh\"ovel, and H.G. Miller, Z. Phys. {\bf 271}, 93(1974).

\bibitem{WeA78}
H.J. Weber and H. Arenh\"ovel, Phys. Rep. {\bf 36}, 279 (1978).

\bibitem{ObP92}
P. Obersteiner, W. Plessas, and J. Pauschenwein, 
Few-Body Syst. Suppl. {\bf 5}, 140 (1992).

\bibitem{Mac99}
R.\ Machleidt, nucl-th/9909036. 

\end{references}
\end{document}